# Emergence of Deviance Detection in Cortical Cultures through Maturation, Criticality, and Early Experience

Zhuo Zhang[1], Amit Yaron[2], Dai Akita[1], Tomoyo Isoguchi Shiramatsu[1], Zenas C. Chao[2], Hirokazu Takahashi[1,2*]

[1]Department of Mechano-Informatics, Graduate School of Information Science and Technology, The University of Tokyo, Tokyo, Japan

[2]International Research Center for Neurointelligence (WPI-IRCN), UTIAS, The University of Tokyo, Tokyo, Japan

**\* Correspondence:**
Hirokazu Takahashi
takahashi@i.u-tokyo.ac.jp

**Abstract**

Mismatch negativity (MMN) in humans reflects deviance detection (DD), a core neural mechanism of predictive processing. However, the fundamental principles by which DD emerges and matures during early cortical development—potentially providing a neuronal scaffold for MMN—remain unclear. Here, we tracked the development of DD in dissociated cortical cultures grown on high-density CMOS microelectrode arrays from 10 to 35 days in vitro (DIV). Cultures were stimulated with oddball and many-standards control paradigms while spontaneous and evoked activity were recorded longitudinally. At early stages, stimulus-evoked responses were confined to fast components reflecting direct activation. From DIV15-20 onward, robust late responses appeared, and deviant stimuli progressively evoked stronger responses than frequent and control stimuli, marking the onset of DD. By DIV30, responses became stronger, faster, and more temporally precise. Neuronal avalanche analysis revealed a gradual transition from subcritical to near-critical dynamics, with cultures exhibiting power-law statistics showing the strongest deviant responses. Nonetheless, DD was also present in non-critical networks, indicating that criticality is not required for its emergence but instead stabilizes and amplifies predictive processing as networks mature. Early oddball experience reinforces the deviant pathway, resulting in faster conduction along those circuits. However, as frequent and deviant pathways become less distinct, the deviance detection index is reduced. Together, these findings demonstrate that DD arises intrinsically through local circuit maturation, while self-organization toward criticality and early experience further refine its strength and timing, providing mechanistic insight into predictive coding in simplified cortical networks and informing the design of adaptive, prediction-sensitive artificial systems.

## 1 Introduction

The ability of neural systems to detect deviations from predicted regularities is a fundamental feature of predictive processing. In humans, this function is indexed by the mismatch negativity (MMN), a prominent event-related potential peaking at ∼100–250 ms after the onset of a deviant stimulus (Näätänen et al., 2007). Comparable mismatch responses (MMRs) have been reported in animal models, typically appearing with shorter latencies of 50–150 ms (Javitt et al., 1992; Shiramatsu et al., 2013). Two main processes contribute to these population-level signals: stimulus-specific adaptation (SSA) and deviance detection (DD). SSA refers to a reduction in neural responses to repetitive "standard" stimuli that does not, or only partially, generalize to infrequent "deviant" stimuli (Movshon and Lennie, 1979). SSA has been observed across multiple stages of the auditory pathway, including the brainstem, inferior colliculus, thalamus, and auditory cortex (Malmierca et al., 2009; Anderson et al., 2009; Duque et al., 2012, 2018; Anderson and Malmierca, 2013; Malmierca et al., 2015). By contrast, DD represents a higher-order computation in which a subset of cortical neurons selectively encodes violations of expected sensory patterns, often manifested as enhanced firing responses to deviant inputs (Shiramatsu et al., 2013; Hamm and Yuste, 2016; Musall et al., 2017; Parras et al., 2017; Polterovich et al., 2018; Hamm et al., 2021). Importantly, DD depends on N-methyl-d-aspartate (NMDA) receptor–mediated synaptic transmission (Shiramatsu et al., 2013; Chen et al., 2015; Zhang et al., 2025).

Developmental studies in humans and animals have shown that mismatch responses emerge early but undergo substantial refinement with maturation. In newborns, deviant stimuli evoke broad positive slow waves rather than adult-like negative deflections, indicating an immature form of change sensitivity (Sambeth et al., 2009; Vestergaard et al., 2010). By 2–4 months of age, infants begin to transition from these positive responses to MMN-like negativities, reflecting the rapid maturation of mechanisms for change detection (Trainor et al., 2003; He et al., 2009). During later childhood, MMN and P3a remain relatively stable, but late discriminative components decline in amplitude, while oscillatory activity in the delta and theta bands strengthens in parallel with the development of attentional control (Morales et al., 2023). Animal studies further show that SSA follows a sequential maturation along the auditory pathway—emerging first in the inferior colliculus, then in the thalamus, and finally in the auditory cortex—where cortical maturation is highly experience dependent (Valerio et al., 2024). Together, these findings suggest that SSA arises early and largely reflects local mechanisms, whereas true DD requires further cortical maturation.

In addition to developmental changes, DD is strongly shaped by sensory experience. Early postnatal noise exposure disrupts the normal maturation of SSA in auditory cortex neurons, whereas the same manipulation in adulthood has little effect (Wang et al., 2019). Conversely, even in mature animals, both passive exposure to specific sounds and the use of conditioned stimulus in classical conditioning can induce long-term memory traces and modify mismatch responses (Kurkela et al., 2016; Shiramatsu et al., 2018, 2021). Consistent with these in vivo findings, in vitro preparations also exhibit experience-dependent predictive properties. For example, dissociated cortical neurons can learn and classify stimulation patterns through training-induced structural reorganization (Shao et al., 2025), and organotypic slices trained with temporal sequences can replay learned patterns during spontaneous activity and generate prediction error-like responses when expected stimuli are omitted (Liu and Buonomano, 2025). Together, these results highlight that experience plays a critical role in shaping predictive processing across developmental stages, both in vivo and in vitro.

Cultured cortical networks grown on multi-electrode arrays provide a powerful reductionist model to investigate how neural dynamics and computations emerge from local interactions (Yaron et al., 2025). During in vitro development, dissociated networks self-organize from sparse, uncoordinated firing into structured activity characterized by synchronous bursts and propagating "network spikes," reflecting the formation of hierarchical connectivity (Jimbo et al., 1999; Eytan and Marom, 2006; Marom and Shahaf, 2002). A hallmark of this process is the emergence of neuronal avalanches whose size distributions approach power laws, indicating that networks operate near a critical regime (Beggs and Plenz, 2003; Pasquale et al., 2008; Priesemann et al., 2013). Transitions toward criticality depend on excitation-inhibition balance, small-world topology, and plasticity mechanisms (Massobrio et al., 2015; Yada et al., 2017; Wilting and Priesemann, 2019; Stepp et al., 2015; Ikeda et al., 2023, 2025), with cultures often residing in slightly subcritical "reverberating" states (Priesemann et al., 2014; Heiney et al., 2022). Importantly, networks approaching criticality exhibit enhanced dynamic range, efficient information transmission, and superior learning performance (Kinouchi and Copelli, 2006; Haldeman and Beggs, 2005; Shew et al., 2009, 2011; Shew and Plenz, 2013; Habibollahi et al., 2023). Thus, criticality emerges as both a developmental outcome of self-organization and a functional regime that may facilitate higher-order computations such as DD.

Building on these insights, high-density CMOS-based MEAs enable simultaneous, long-term recordings from thousands of neurons with subcellular resolution, providing unprecedented access to the spatiotemporal structure of cultured network activity (Berdondini et al., 2005; Müller et al., 2015; Obien et al., 2015; Bakkum et al., 2013, 2019; Yaron et al., 2025; Yada et al., 2017). Recent studies using these platforms have demonstrated that predictive computations also emerge in cultured neurons. Early stimulus-evoked responses (0–10 ms) are dominated by direct activation, whereas later responses (11–150 ms) reflect synaptic recruitment (Eytan et al., 2003), giving rise to SSA- and DD-like phenomena analogous to mismatch responses observed in vivo (Zhang et al., 2025). However, it remains unclear how DD develops over time in vitro, and whether its maturation is mechanistically linked to the establishment of critical network dynamics. In this study, we systematically tracked the developmental trajectory of DD in dissociated cortical cultures between 10 and 35 days in vitro (DIV10 and DIV 35), leveraging high-density CMOS-MEA recordings to monitor spontaneous and evoked activity longitudinally. We further tested whether cultures operating closer to criticality exhibit stronger DD, and whether early structured stimulation modifies the temporal dynamics of predictive processing at maturity. This approach allowed us to directly probe how self-organized network dynamics support the emergence and refinement of prediction error processing in simplified cortical circuits.

## 2   Methods

We conducted experiments on dissociated cortical cultures recorded with a high-density CMOS-MEA between DIV10 and DIV35. The study consisted of two complementary parts. First, we tracked the developmental trajectories of DD and network dynamics. DD was assessed by repeating oddball and many-standards control (MSC) paradigms every five days from DIV10 to DIV30, while spontaneous network activity was analyzed to quantify neuronal avalanches and classify network states based on their avalanche size distributions: exponential (subcritical), bimodal (sporadic large avalanches coexisting with small ones), or power-law (near-critical). Second, we tested the lasting effects of early structured stimulation. One group of cultures received the DIV10–30 stimulation protocol, whereas a control group remained unstimulated; both groups were then assessed at maturity (DIV31–35) under the same paradigms.

### 2.1 Preparation of Neuronal Culture

Primary cortical neurons were obtained from embryonic day 18 (E18) Wistar rats of both sexes. All procedures were approved by the Animal Experiments Committee of the Graduate School of Information Science and Technology, the University of Tokyo (JA21-8), and were conducted in accordance with the guidelines of the Japanese Physiological Society.

Pregnant rats were euthanized under deep anesthesia, and cortical tissues were dissected from both hemispheres of E18 fetuses. Following enzymatic digestion with 0.25% trypsin-EDTA (Thermo Fisher Scientific, Waltham, MA, USA), cells were gently dissociated in Neurobasal Plus medium supplemented with B-27 Plus (Thermo Fisher Scientific) and adjusted to a density of 38,000 cells per 5 μL. A 5-μL suspension was seeded onto high-density CMOS microelectrode arrays (MaxOne, MaxWell Biosystems, Switzerland) pre-coated with 0.05% polyethyleneimine (PEI; Sigma-Aldrich, St. Louis, MO, USA) and laminin (Sigma-Aldrich). After a 2-h adhesion period, cultures were covered with 0.6 mL of Neurobasal Plus medium (Figure 1A). Cells were incubated at 36.5 °C in a humidified atmosphere containing 5% $CO_2$. On the following day, half of the medium was replaced with growth medium consisting of DMEM (Thermo Fisher Scientific), horse serum (Cytiva, Marlborough, MA, USA), GlutaMAX (Thermo Fisher Scientific), and sodium pyruvate (Thermo Fisher Scientific), as described previously (Jimbo et al., 1993; Potter et al., 2001; Yada et al., 2016, 2017; Kubota et al., 2019; Ikeda et al., 2025; Zhang et al., 2025). Thereafter, half-medium changes were performed every 3–4 days. Recordings were conducted between DIV10 and DIV35 to capture the developmental trajectory of activity.

## 2.2 Electrophysiological Stimulation and Recording

Electrophysiological experiments were performed using a high-density CMOS microelectrode array system (MaxOne, MaxWell Biosystems, Switzerland) comprising 26,400 electrodes arranged in a 220 × 120 grid over a 3.85 × 2.1 mm² area. Each electrode measured 9.3 × 5.45 μm with an inter-electrode spacing of 17.5 μm, and up to 1,024 electrodes could be simultaneously selected for high-resolution extracellular recordings. Signals were sampled at 20 kHz.

Because neurons were randomly distributed across the array, spontaneous activity was first assessed by scanning overlapping electrode patches to generate a spatial activity map. Based on this map, 1,024 electrodes in highly active regions, preferentially near dense neuronal clusters, were selected for subsequent recordings. Continuous 30-min recordings from these electrodes were used to characterize spontaneous activity, including spike amplitudes and firing-rate distributions.

Stimulation electrodes were identified from the same activity map. Electrodes located near putative excitatory and inhibitory neurons—classified on the basis of relative spike timing and waveform characteristics—were examined, and sites associated with excitatory activity were chosen for stimulation (Tajima et al., 2015; Ikeda et al., 2025). Six electrodes within the same active cluster were stimulated simultaneously, with at least one electrode spacing between them to minimize cross-activation. Biphasic voltage pulses (200 μs positive phase followed by 200 μs negative phase, 350 mV amplitude) were delivered to each stimulation site.

## 2.3 Experimental Paradigms

To investigate the developmental trajectory of DD in dissociated cortical cultures, two structured stimulation experiments were conducted.

In the first experiment, DD was examined across immature to mature developmental stages (DIV10–30). Each culture underwent oddball stimulation every five days (Figure 1B), resulting in five sessions in total. For each session, a 30-minute baseline recording of spontaneous activity was obtained before stimulation, followed by a 30-minute post-stimulation recording to assess subsequent network dynamics. Stimulation protocols included both the oddball and many-standard control (MSC) paradigms, described below.

In the second experiment, we tested whether early-life exposure to structured stimulation influenced DD at later stages. Cultures were divided into two groups: a stimulated group that received the five-session protocol during DIV10–30, and a naïve control group that remained unstimulated during this period (Figure 1C). Both groups were then subjected to the same oddball and MSC paradigms in the mature phase (DIV31–35) to assess whether early stimulation modulated DD expression in mature networks.

Across both experiments, oddball and MSC stimulation paradigms were applied. In the oddball paradigm, three pairs of standard and deviant sites were selected from ten predefined stimulation locations (Stim A–J). Stimuli were presented in a pseudo-randomized sequence with a 9:1 standard-to-deviant ratio. A flip-flop design was used such that standard and deviant roles were reversed across two successive runs, allowing within-site comparisons under frequent versus rare presentation. Each run consisted of 600 stimuli delivered at 500-ms intervals, with a 5-min rest period between runs. In the MSC paradigm, all ten sites (Stim A–J) were presented in a randomized sequence with equal probability (10% each). This condition provided a non-regular baseline, enabling the separation of true DD from repetition suppression or adaptation. Each MSC run also consisted of 600 stimuli delivered at 500 ms inter-stimulus intervals (ISIs). The stimulus sequence for both paradigms is illustrated in Figure 1B.

**2.4 Data preprocessing and deviance detection analysis**

Extracellular signals were band-pass filtered between 300 and 3000 Hz for spike detection. To minimize stimulation artifacts, the first 1 ms following each stimulus was excluded from the detection window. For each paradigm, three stimulation site-groups were tested to reduce potential bias in site preference, and the group that exhibited the largest evoked response was selected for subsequent analyses. Stimulus-evoked firing rates were calculated from peri-stimulus time histograms (PSTHs) constructed with 1-ms bins.

Neuronal responses were analyzed under three conditions: standard, deviant, and MSC. The MSC condition was designed so that stimuli occurred with the same overall frequency as the deviant stimuli but were distributed across at multiple locations, thereby serving as a control to distinguish true DD from SA.

SSA was quantified using a SSA index (SSAI):

$$SI(R) = \frac{R - R_{std}}{R + R_{std}},$$

where $R$ represents the response to deviant ($R_{dev}$) or MSC ($R_{MSC}$) stimuli, and $R_{std}$ is the response to standard stimuli. A positive SSAI indicates that responses to deviant or MSC stimuli exceeded those to standards, while a greater SSAI for deviant compared with MSC reflects true DD beyond adaptation.

To directly compare deviant and MSC responses, we further defined a deviance detection index (DDI):

$$DDI = \frac{R_{dev} - R_{MSC}}{R_{dev} + R_{MSC}}.$$

A positive DDI indicates stronger responses to deviant than to MSC stimuli, providing a direct measure of true DD. Both SSAI and DDI were quantified specifically from the late response window (11–150 ms post-stimulus), as this period has been demonstrated to predominantly reflect synaptic transmission–driven network activity (Zhang et al., 2025).

In addition to SSAI and DDI, we also quantified temporal features of the evoked responses. The response duration was defined as the post-stimulus period during which the response amplitude exceeded three standard deviations above the mean baseline activity (50 ms preceding stimulus onset). The peak latency was defined as the time from stimulus onset to the maximum of the population-averaged PSTH within 0–200 ms. For each culture, PSTHs were first computed channel-wise and then averaged across responsive channels to obtain the population response.

All analyses were performed at the culture level. Unless otherwise specified, each data point corresponds to one independent culture (N = 7 cultures across DIV10–30, with multiple recordings per culture). For mature-stage comparisons (DIV31–35), N = 6 cultures were analyzed in each group (experimental vs. control).

## 2.5 Avalanche definition

Neuronal avalanches were defined as spatiotemporal clusters of spikes identified through temporal contiguity. All spikes detected across electrodes were merged into a single time-ordered sequence ("array-wide" spike train). If the interval between two consecutive spikes was shorter than a threshold duration $\Delta t$, those spikes were assigned to the same avalanche. If no subsequent spike occurred within $\Delta t$, the preceding spike was considered an avalanche of size one (single-spike event). In this way, each spike belonged to exactly one avalanche, and avalanches are separated by silent periods larger than $\Delta t$. The threshold $\Delta t$ was determined according to the intrinsic activity timescale: on each recording day, $\Delta t$ was set to the mean inter-spike interval of the network, calculated across all active electrodes. The method follows established approaches (Pasquale et al., 2008; Tetzlaff et al., 2010; Yada et al., 2017) and does not rely on a fixed bin size, thereby avoiding bias introduced by arbitrary temporal discretization.

## 2.6 Characterization of avalanche distributions

For each recording, we analyzed the probability distribution of avalanche sizes $s$ (number of spikes per avalanche). Two candidate models were fitted: a power-law distribution and an exponential distribution (Yada et al., 2017). The probability mass functions were defined as:

$$P_\alpha(s) = \begin{cases} cs^{-\alpha}, & s_{min} \leq s \leq s_{max} \\ 0, & otherwise \end{cases},$$

$$P_\lambda(s) = \begin{cases} ce^{-\lambda s}, & s_{min} \leq s \leq s_{max} \\ 0, & otherwise \end{cases},$$

where $c$ is a normalization constant.

The fitting range of the power-law was defined from the minimum cutoff $s_{min} = 2$ (excluding single-spike avalanches) up to the maximum avalanche size observed ($s_{max}$). The exponential distribution was fitted across the entire size range. Because very small avalanches (particularly size 1) often deviate from ideal scaling, our power-law fits primarily considered avalanches with $s > 2$ ($s_{min} = 2$).

The power-law exponent $\alpha$ and exponential rate parameter $\lambda$ were estimated using maximum likelihood estimation (MLE). For a dataset of avalanche sizes $\{s_i\}_{i=1}^{N}$, the log-likelihood functions were:

$$logL_{pl}(\alpha) = \sum_{i=1}^{N} logP_\alpha(s_i),$$

$$logL_{ep}(\lambda) = \sum_{i=1}^{N} logP_\lambda(s_i).$$

Goodness-of-fit was assessed using the Kolmogorov–Smirnov (KS) statistic:

$$D = \max_s |F_{emp}(s) - F_{model}(s)|,$$

where F(s) denotes the cumulative distribution function (CDF). A lower KS statistic indicates a closer fit. To compare relative model support, we calculated the log-likelihood ratio (LLR):

$$LLR = logL_{pl} - logL_{ep}.$$

and the Akaike Information Criterion (AIC):

$$AIC = -2logL + 2k,$$

where $k = 1$ is the number of free parameters. The model with the lower AIC was considered better supported, unless $\Delta AIC < 2$, in which case the models were regarded as indistinguishable.

These statistical measures were jointly used to determine whether avalanche size distributions were better described by a power-law scaling or an exponential form.

**2.7 Classification criteria**

Avalanche size distributions were classified into power-law, exponential, or bimodal types (Yada et al., 2017).

First, bimodality was assessed by examining the largest gap in avalanche sizes on a logarithmic scale. The maximum difference between consecutive values in $log_{10}(s)$ was defined as the gap value $G$. If $G$ exceeded the threshold defined as the third quartile (Q3) of all gap values across the dataset, the distribution was labeled as bimodal, reflecting the presence of unusually large avalanches separated from the main body of smaller events.

For non-bimodal distributions, we compared the fits of power-law and exponential models.

- A distribution was classified as power-law if the log-likelihood ratio favored the power-law fit ($LLR > 0$) and the KS statistic for the power-law fit was below the acceptance threshold.
- Conversely, a distribution was classified as exponential if $LLR < 0$ and the exponential fit satisfied the KS criterion.

When both models provided comparable fits, additional criteria were applied. Specifically, when the difference in $AIC$ ($\Delta AIC$) was smaller than the standard cutoff ($|\Delta AIC| < 2$), the ratio of KS statistics (acceptance tolerance) was used to resolve the classification, assigning the distribution to either power-law or exponential. Using this combination of gap-based and statistical criteria (KS, LLR, $\Delta AIC$), each avalanche distribution was ultimately classified as power-law, exponential, bimodal.

**2.8 Statistical analysis**

All statistical analyses were performed in Python and are reported as mean ± SEM unless otherwise specified. Normality was assessed using the Shapiro–Wilk test. SSAI for deviant and MSC conditions were each tested against zero using two-tailed Wilcoxon signed-rank tests. Direct comparisons between deviant and MSC responses were performed using paired Wilcoxon signed-rank tests, and DDI values were similarly tested against zero.

Group comparisons across developmental stages or avalanche categories were assessed with the Kruskal–Wallis test, followed by Dunn's post hoc tests with Holm correction. Where assumptions of normality were met, parametric analyses including two-way repeated-measures ANOVA,were applied, and effect sizes were reported as partial η². For nonparametric tests, effect sizes were estimated using Cliff's delta (δ).

Multivariate group differences were further examined using permutational multivariate analysis of variance (PERMANOVA) with 999 permutations. False discovery rate (FDR) correction (Benjamini–Hochberg method) was applied to adjust for multiple comparisons across univariate tests. Statistical significance was defined as $p < 0.05$ after correction.

## 3 Results

### 3.1 Developmental Emergence of Deviance Detection (DIV10–30)

#### 3.1.1 Emergence and stabilization of DD across development

We first investigated when and how DD (deviance detection) emerges and evolves in vitro by analyzing neuronal responses to standard, deviant, and MSC stimuli across five developmental stages (DIV10, 15, 20, 25, and 30).

At DIV10, dissociated cortical cultures already exhibited spontaneous network activity, including synchronous bursts (Figure 2A-B). However, stimulus-evoked responses were largely restricted to early components, reflecting direct electrical activation (Figure 3A). Clear late responses (11–150 ms post-stimulus)—typically associated with synaptically mediated network recruitment (Jimbo et al., 2000; Eytan and Marom, 2006; Bakkum et al., 2008; Kermany et al., 2010; Gal et al., 2010; Dranias et al., 2013; Zhang et al., 2025)—were absent at this stage. From DIV15 onward, stimulation consistently evoked late responses, whose amplitudes increased progressively with maturation (Figure 3B–E).

With respect to SSA, the SSAI for deviant responses became significantly greater than zero from DIV15 onward, indicating stronger late responses to deviants compared with standards ($p = 0.078$ at DIV10, $p < 0.05$ at later stages; one-sample Wilcoxon signed-rank test, two-tailed). From DIV20, deviant responses also exceeded MSC responses ($p = 0.688$ at DIV10, $p = 0.078$ at DIV15, $p < 0.05$ at later stages, Wilcoxon signed-rank test; Figure 3F-J). Analyses of the DDI further showed that values were not significantly above zero at DIV10 ($p = 0.078$; Figure 4A) but became significant from DIV15 onward (all $p < 0.05$). Although small non-zero DDI values were obtained at DIV10, these likely reflect weak fluctuations within the 11–150 ms window rather than deviance detection, as no robust late responses were present at this stage (Figure 3A). DD peaked at DIV15–20 and remained elevated through DIV25–30, indicating that DD emerges around DIV15 and stabilizes with maturation. Notably, SSAI and DDI yielded slightly different results at DIV15, reflecting their distinct definitions: SSAI compares deviant or MSC responses with standards, whereas DDI directly contrasts deviant vs MSC. From DIV20 onward, both measures converged, consistently indicating robust DD. These findings suggest that true DD—beyond simple adaptation—emerges around DIV20, accompanying the gradual establishment of predictive processing.

In parallel, peak latencies shifted progressively earlier between DIV15 and DIV30 not only for deviant responses (Figure 4B) but also for standard and MSC, reflecting a general developmental acceleration of evoked responses. Consistently, response duration decreased with maturation (Table 1), indicating improved temporal precision. Spontaneous firing rates (Figure 4C) and evoked response amplitudes (Figure 4D) both increased over development, reflecting enhanced baseline excitability and sensory responsiveness. A two-way ANOVA confirmed significant main effects of DIV ($F = 30.1$, $p < 0.0001$, partial $\eta^2 = 0.57$) and Paradigm ($F = 27.8$, $p < 0.0001$, partial $\eta^2 = 0.38$), but no interaction ($p = 0.21$). Thus, firing rates increased with maturation and differed across paradigms, with deviant stimuli generally eliciting stronger activity than standards. Although Figure 4B illustrates peak latencies for deviant responses, analyses of standard and MSC showed similar developmental decreases (Table 1), indicating that all response types became faster with maturation.

Together, these findings demonstrate that as dissociated cortical networks develop from DIV10 to DIV30, stimulus-evoked responses become progressively stronger, faster, and more temporally precise. In parallel, DD capacities strengthen and stabilize, reaching a mature state by DIV30.

**3.1.2 Relationship between network criticality stage and DD**

We next investigated whether and how the development of DD is associated with the maturation of network dynamics. As dissociated cortical networks mature, their spontaneous activity progressively shifts from a subcritical to a near-critical state (Tetzlaff et al., 2010; Yada et al., 2017). This transition reflects ongoing synaptic wiring and rewiring, during which GABA-mediated giant depolarizing potentials facilitate the integration of immature neurons (Sipilä, 2005; Le Magueresse and Monyer, 2013). Building on these observations, we hypothesized that the criticality state of the network influences both the development and expression of DD. To test this, we examined the relationship between avalanche dynamics and deviance-related responses across developmental stages.

The avalanche size distributions of cultures at different DIVs exhibited a progressive shift from exponential to power-law scaling (Figure 5A–C, Figure 2). Exponential distributions predominated at DIV10, consistent with immature, subcritical dynamics (Pasquale et al., 2008; Yada et al., 2017), but gradually declined with further development. Bimodal distributions, characterized by sporadic large avalanches, were mainly observed at DIV10 and DIV15. Power-law distributions first appeared around DIV15 and become the dominant form after DIV20 (Figure 5D).

We next examined whether DD differed across cultures classified as exponential, bimodal, or power-law. The DDI showed a stepwise increase across these categories (Figure 5E). A Kruskal–Wallis test revealed a significant effect of distribution type on DDI (H = 13.36, p = 0.0013). Post-hoc Dunn's tests with Holm adjustment indicated that cultures with power-law distributions exhibited significantly higher DDI values than those with exponential distributions (p = 0.0008), whereas bimodal cultures did not differ significantly from either group. Effect size estimation using Cliff's delta confirmed a large difference between exponential and power-law cultures ($\delta = -0.91$), supporting the robustness of this contrast.

To test whether the observed relationship between avalanche distribution and DDI could be explained by overall network excitability, we compared firing rates across exponential, bimodal, and power-law groups (Figure 5F). The Kruskal–Wallis tests revealed no significant differences between groups (H = 2.45, p = 0.29, $\varepsilon^2$ = 0.014), and post-hoc Dunn's tests with Holm correction confirmed the absence of pairwise differences (all p > 0.6). Effect sizes were small to negligible (Cliff's $\delta$ range –0.36 to –0.03). Correlation analysis further indicated that DDI was not associated with firing rate (Spearman $\rho$ = 0.03, p = 0.85, 95% CI [–0.27, 0.35]; Figure 5G). These results demonstrate that the association between avalanche category and DD cannot be explained by firing rate differences.

Together, these findings suggest that cultures whose avalanche statistics approximate power-law scaling tend to exhibit stronger DD, whereas DD is already present in cultures with bimodal avalanche distributions. Thus, near-critical dynamics are not required for the emergence of DD but appear to reinforce and stabilize deviance processing as networks mature.

**3.2 Lasting Effects of Early Structured Stimulation (assessed at DIV31–35)**

We finally evaluated whether early sensory stimulation influences the maturation of DD. At DIV31–35, we compared two groups of cultured cortical networks in the mature stage: an experimental group that had undergone five structured stimulation sessions during the immature period (DIV10–30), and a control group without prior stimulation.

In both groups, neurons exhibited robust deviant responses under the oddball paradigm (Figure 6A, D). Time-resolved SSAI analyses further demonstrated DD in both groups (Figure 6B, E), indicating that the fundamental mechanisms of predictive processing emerged in vitro regardless of early sensory experience. However, only the control group showed significantly greater SSAI responses to deviant stimuli compared with the MSC condition (p = 0.031, Wilcoxon signed-rank test; Figure 6C), whereas the experimental group did not reach significance (p = 0.063; Figure 6F). Nevertheless, in both groups, the SSAI values for deviant stimuli were significantly greater than zero (p < 0.05, one-sample Wilcoxon signed-rank test, two-tailed), confirming the presence of SSA. Consistently, both groups also exhibited DDI values significantly greater than zero (p = 0.031 for the experimental group, p = 0.016 for the control group; Figure 6G), demonstrating that DD was present in both groups, albeit stronger in controls.

PERMANOVA revealed a significant overall difference between experimental and control groups (pseudo-F = 6.81, p = 0.008, 999 permutations). Subsequent univariate tests with FDR correction identified group differences across several dimensions of deviant processing: DDI was higher in controls (p_FDR = 0.032; Figure 6G, H), whereas the experimental group exhibited shorter peak latencies (p_FDR = 0.032; Figure 6I), shorter response durations (p_FDR = 0.036; Table 2), and higher firing rates (p_FDR = 0.018; Figure 6J).

Together, these results suggest that early exposure to structured sensory stimulation induces lasting modifications in network organization, likely through long-term synaptic plasticity. Stimulated cultures exhibited shorter response durations but lower DDI, indicating responses that were faster yet less differentiated. These findings highlight the capacity of developing cortical networks to integrate early experience into their functional architecture, thereby shaping the temporal dynamics of predictive processing even in the absence of continued sensory input.

## 4 Discussion

This study systematically examined the developmental trajectory, network dynamics, and experience-dependent modulation of DD in dissociated cortical networks. We found that DD strengthened progressively during maturation: SSA emerged by DIV15, and robust DD stabilized by DIV30, accompanied by faster and more temporally precise network responses. At the population level, avalanche analysis revealed that cultures approaching criticality (power-law statistics) tended to exhibit stronger DD, whereas subcritical (exponential) cultures showed weaker responses. Importantly, DD was already present in cultures with bimodal avalanche distributions, indicating that critical dynamics are not strictly required for its emergence. Instead, criticality appears to reinforce and stabilize DD once networks mature. Furthermore, early structured stimulation produced lasting modifications at maturity, resulting in faster but less differentiated processing. Together, these findings suggest that DD is an intrinsic property of developing cortical networks, further refined by both network self-organization toward criticality and early sensory experience, at least in simplified in vitro systems.

**4.1 Self-organized maturation of deviance detection**

Local cortical slices have been shown to learn temporal information through optical stimulation and to exhibit predictive capabilities independent of top-down inputs (Liu and Buonomano, 2025). Similarly, dissociated cortical neurons derived from rat embryos display sensitivity to both deviance and regularity, reflecting an intrinsic capacity for predictive processing (Zhang et al., 2025). Our findings further support the view that DD can develop spontaneously in dissociated cortical networks, even in the absence of structured sensory input.

SSA emerged around DIV15 and became progressively more stable and efficient with further development, as indicated by increasingly differentiated responses to deviant versus standard stimuli and progressively shorter response responses. This developmental trajectory differs from that observed in the rat auditory cortex (A1), where SSA matures gradually with a decrease in common SSAI (Valerio et al., 2024). DD, a higher-level process than adaptation, also appeared around DIV15-20, but full maturation and stabilization were not achieved until DIV30.

These results indicate a staged, hierarchical progression in which adaptation to regular input and sensitivity to deviance provide the foundation for building predictive models. This progression underscores the intrinsic, self-organizing capacity of neuronal networks to optimize information processing, consistent with theories of predictive coding and efficient coding. The developmental shortening of response durations and peak latencies of deviance further supports that information processing becomes faster and more efficient in mature networks, analogous to the decrease in mismatch response latency observed in vivo as neural circuits refine (Näätänen et al., 2019).

In summary, our findings demonstrate that DD and SSA do not require complex hierarchical circuits or structured inputs but instead emerge as natural products of intrinsic network dynamics, providing an experimentally tractable model for studying predictive computation (Mill et al., 2011; Kern and Chao, 2023).

**4.2 Critical dynamics as a substrate for deviance detection**

Numerous studies have shown that dissociated neural networks gradually transition from subcritical to critical dynamics during development (Kamioka et al., 1996; Pasquale et al., 2008; Sun et al., 2010; Tetzlaff et al., 2010; Yada et al., 2017). Consistent with this, we found that early in vitro activity exhibited exponentially distributed avalanche sizes, indicative of a subcritical regime. As synaptic connectivity matured, cascades grew larger and more frequent, yielding bimodal avalanche size distributions characterized by many small events interspersed with occasional large bursts. These may reflect the predominance of excitatory synapses, during early culture stages (Chiappalone et al., 2006). By DIV20–30, mature cultures displayed power-law avalanche size distributions—a hallmark of self-organized criticality (SOC), in which event sizes span a broad range without a characteristic scale. As development progressed, inhibitory synapses strengthened, possibly through GABA-mediated mechanism, allowing networks to self-organize toward a balanced state (Baho and Di Cristo, 2012; Le Magueresse and Monyer, 2013; Plenz et al., 2021). This shift fragmented large burst into more controlled propagation (Yada et al., 2017).

Notably, cultures approaching a power-law distribution showed the strongest DD, whereas subcritical cultures (exponential distributions) exhibited much weaker deviance sensitivity, intermediate bimodal networks displayed only modest DD. These results suggest that near-critical dynamics facilitate predictive processing. Importantly, this relationship was not explained by differences in overall network excitability: average firing rates did not differ significantly across avalanche categories, and DDI values were uncorrelated with firing rate. Thus, the qualitative network state (critical vs. subcritical), rather than absolute activity level, determines the efficacy of deviance processing.

This observation aligns with theoretical predictions that neural systems operating near criticality maximize dynamic range and information transmission, thereby optimizing the propagation and integration of prediction-error signals (Beggs and Plenz, 2003; Shew et al., 2009, 2011; Shew and Plenz, 2013; Wilting and Priesemann, 2019). In such networks, small perturbations can propagate widely without destabilizing the system, enabling flexible yet stable encoding of statistical regularities and rapid updating of internal models when deviations occur.

**4.3 Synaptic mechanisms underlying deviance detection**

At the synaptic level, one plausible mechanism for DD in these networks is short-term synaptic adaptation at excitatory synapses for repetitive stimuli, combined with slower NMDA-receptor-mediated currents that amplify responses to rare inputs. Consistent with MMN in intact brains, the late-phase mismatch responses in cultures were found to be NMDA receptor-dependent, implicating NMDA-mediated synaptic transmission and plasticity in DD (Zhang et al., 2025). Notably, DD capability emerged gradually in parallel with the development of synaptic connectivity and network bursts, suggesting that reverberant circuits supporting short-term memory traces may underlie the effect.

In vitro studies of synapse development show that synapse formation is most pronounced during the first 15 days and remains stable up to DIV60. The NMDA receptor subunit NR1 and AMPA receptor subunits GluR2/GluR3 are already expressed by DIV5, preceding the appearance of synaptic markers, and glutamate receptor levels peak by DIV10-15 before stabilizing (Lesuisse and Martin, 2002). Similarly, between 7 and 14 DIV, large increases in the amplitudes and frequencies of spontaneous excitatory postsynaptic currents (EPSCs) have been observed in E18 cultures. EPSCs are primarily mediated by AMPA receptors at DIV7, but by both NMDA and AMPA receptors at 14 DIV (Lin et al., 2002). Moreover, in cortical neurons, the total number of excitatory synapses increases significantly only after DIV21, continuing through DIV28, whereas inhibitory synapses increase significantly only after DIV28 (Harrill et al., 2015).

The developmental trajectory of NMDA-related DD follows a similar course. At DIV10, responses were likely dominated by AMPA receptors, resulting in the absence of slow late responses. By DIV15, NMDA receptor involvement gave rise to prominent late responses, and by DIV20 a relatively stable DD capability was established, corresponding to the gradual maturation and stabilization of synaptic development.

**4.4 Experience-dependent tuning of temporal dynamics**

Although DD was evident in both early-stimulated and control cultures at maturity (DIV31–35), their response profiles diverged. Stimulated cultures exhibited significantly shorter peak latencies to deviant stimuli, but at the cost of slightly lower DDI compared to controls. These findings suggest that early sensory experience does not generate predictive capabilities anew, but rather shapes their developmental trajectory by biasing networks toward faster yet less differentiated deviant detection. Mechanistically, these effects are consistent with experience-dependent synaptic plasticity, including NMDA receptor-mediated long-term potentiation (LTP) and spike-timing-dependent plasticity (STDP). NMDA receptor–mediated LTP strengthens synaptic efficacy (Bliss and Collingridge, 1993), while STDP refines connectivity by reinforcing causally effective, shorter-latency pathways (Bi and Poo, 1998; Song et al., 2000), thereby reducing response latency without necessarily enhancing response contrast. Within predictive coding frameworks, NMDA-dependent plasticity has been proposed to support the refinement of temporal prediction mechanisms (Friston, 2005; Garrido et al., 2009).

Supporting this interpretation, recent in vitro studies demonstrate that repeated stimulation reorganizes excitatory-inhibitory pathways, enhancing pattern recognition (Shao et al., 2025; Ikeda et al., 2025) and enabling spontaneous replay of learned temporal sequences (Liu and Buonomano, 2025). In vivo, passive sensory exposure can induce long-term memory traces even in the adult auditory cortex when stimuli are sufficiently complex (Kurkela et al., 2016), whereas inappropriate early exposure (e.g., noise rearing) disrupts the normal development of SSA and deviance sensitivity (Wang et al., 2019). Similarly, Valerio et al. (2024) showed that both deprivation and overexposure during critical periods distort the maturation of SSA, underscoring the bidirectional plasticity of predictive coding mechanisms. Together, these findings suggest that early structured stimulation reorganizes functional circuitry in a way that prioritizes rapid propagation of prediction error signals, albeit with reduced differentiation between deviant and frequent events.

From a broader perspective, our results highlight how early sensory experience calibrates predictive processing. On the one hand, such tuning may confer efficiency by enabling faster responses to unexpected inputs. On the other hand, excessive or atypical early stimulation may limit sensitivity. Altered MMN amplitude and latency are observed in adolescents with neurodevelopmental disorders such as autism and ADHD (attention deficit hyperactivity disorder) (Chen et al. 2020; Yamamuro et al. 2016; Di Lorenzo et al. 2020; Lassen et al. 2022; Schwartz et al., 2018), potentially linked to abnormal auditory environments during development. Beyond basic neuroscience, these insights also underscore the applied potential of guiding neural computation: controlled early stimulation could be leveraged to steer the functional maturation of both biological and artificial neural systems toward the desired balance of efficiency and sensitivity.

## 5  Conclusion

Our findings demonstrate that deviance detection (DD) in dissociated cortical cultures emerges around DIV15 and stabilizes by DIV30, as stimulus-evoked responses become stronger, faster, and more temporally precise—consistent with the progressive establishment of recurrent synaptic interactions. At the network level, cultures that later exhibited power-law avalanche statistics showed higher DDI than those with exponential distributions. Importantly, DD was already present in cultures with bimodal avalanche distributions and was not significantly different from power-law cultures in our sample, indicating that near-critical dynamics are not required for the emergence of DD. Rather, criticality appears to reinforce and stabilize deviance processing as networks mature.

Early structured stimulation produced lasting changes at maturity (DIV31–35): responses were faster but accompanied by lower DDI compared with unstimulated controls, indicating that early experience biases developing networks toward temporal efficiency at the expense of response differentiation. Together, these results indicate that DD arises intrinsically before near-critical dynamics dominate, while subsequent network self-organization and early experience further refine its strength and timing. This developmental and mechanistic perspective clarifies how predictive processing emerges in cortical circuits and provides inspiration for designing adaptive, prediction-sensitive artificial networks. Moreover, by modeling how DD develops and stabilizes in simplified cortical networks, this work provides a tractable framework for investigating circuit-level dysfunctions that may underlie altered mismatch responses in neurodevelopmental disorders.

**Tables**

Table 1. Response duration across developmental stages. Response duration (ms, mean ± SD) for standard (STD), deviant (DEV), and many-standards control (MSC) stimuli at DIV15, 20, 25, and 30.

| **Days in vitro** | **STD** | **DEV** | **MSC** |
| --- | --- | --- | --- |
| **DIV 15 (ms)** | 225.0±72.4 | 244.7±48.5 | 174.3±78.1 |
| **DIV 20 (ms)** | 198.6±46.4 | 209.4±31.0 | 147.6±47.0 |
| **DIV 25 (ms)** | 162.7±50.3 | 146.0±32.0 | 125.4±29.3 |
| **DIV 30 (ms)** | 160.4±41.8 | 138.1±13.4 | 129.3±27.3 |

Table 2. Response duration in experimental and control groups at maturity. Response duration (ms, mean ± SD) for standard (STD), deviant (DEV), and many-standards control (MSC) stimuli in experimental and control groups during the mature stage (DIV31–35).

| **Groups** | **STD** | **DEV** | **MSC** |
|---|---|---|---|
| **Experimental group (ms)** | 155.3±39.0 | 139.7±18.6 | 144.2±49.6 |
| **Control group (ms)** | 173.8±65.9 | 160.5±33.3 | 118.2±48.2 |

**Figures**

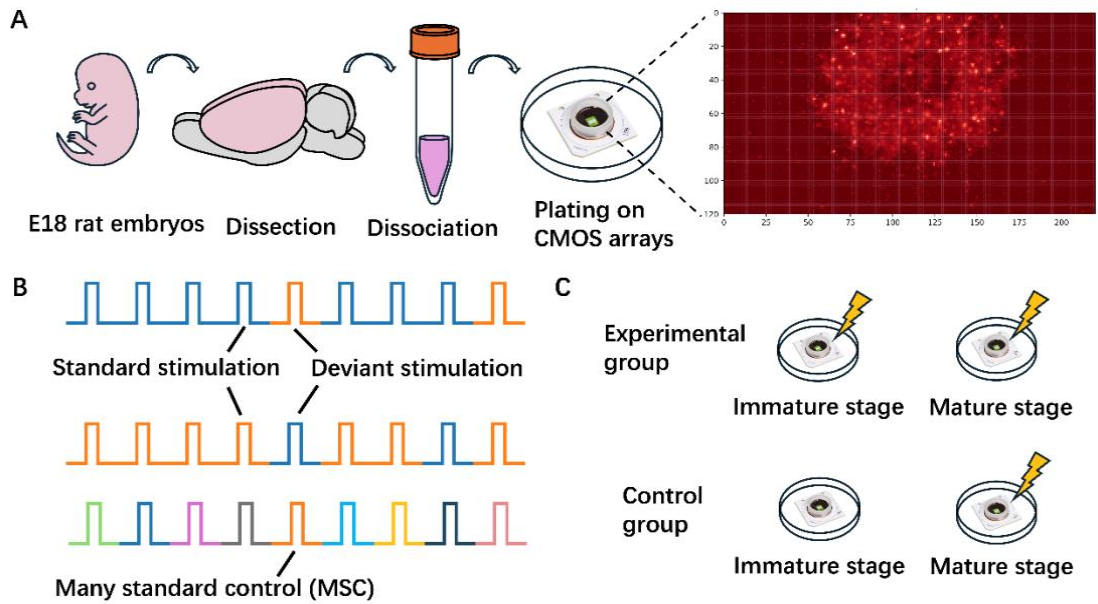

Figure 1. Experimental procedure. (A) Culture preparation. Cortical tissues were dissected from embryonic day 18 (E18) rat brains, enzymatically and mechanically dissociated, and plated onto high-density CMOS microelectrode arrays. Neurons attached randomly across the electrode surface. (B) Stimulation paradigms. Oddball and many-standards control (MSC) paradigms were applied with a fixed inter-stimulus interval of 500 ms. (C) Early exposure experiment. To assess the impact of early sensory stimulation, cultures were divided into two groups. The experimental group received oddball stimulation every five days during the immature period (DIV10–30), whereas control cultures remained unstimulated. At the mature stage (DIV31–35), both groups were tested using the oddball and MSC paradigms.

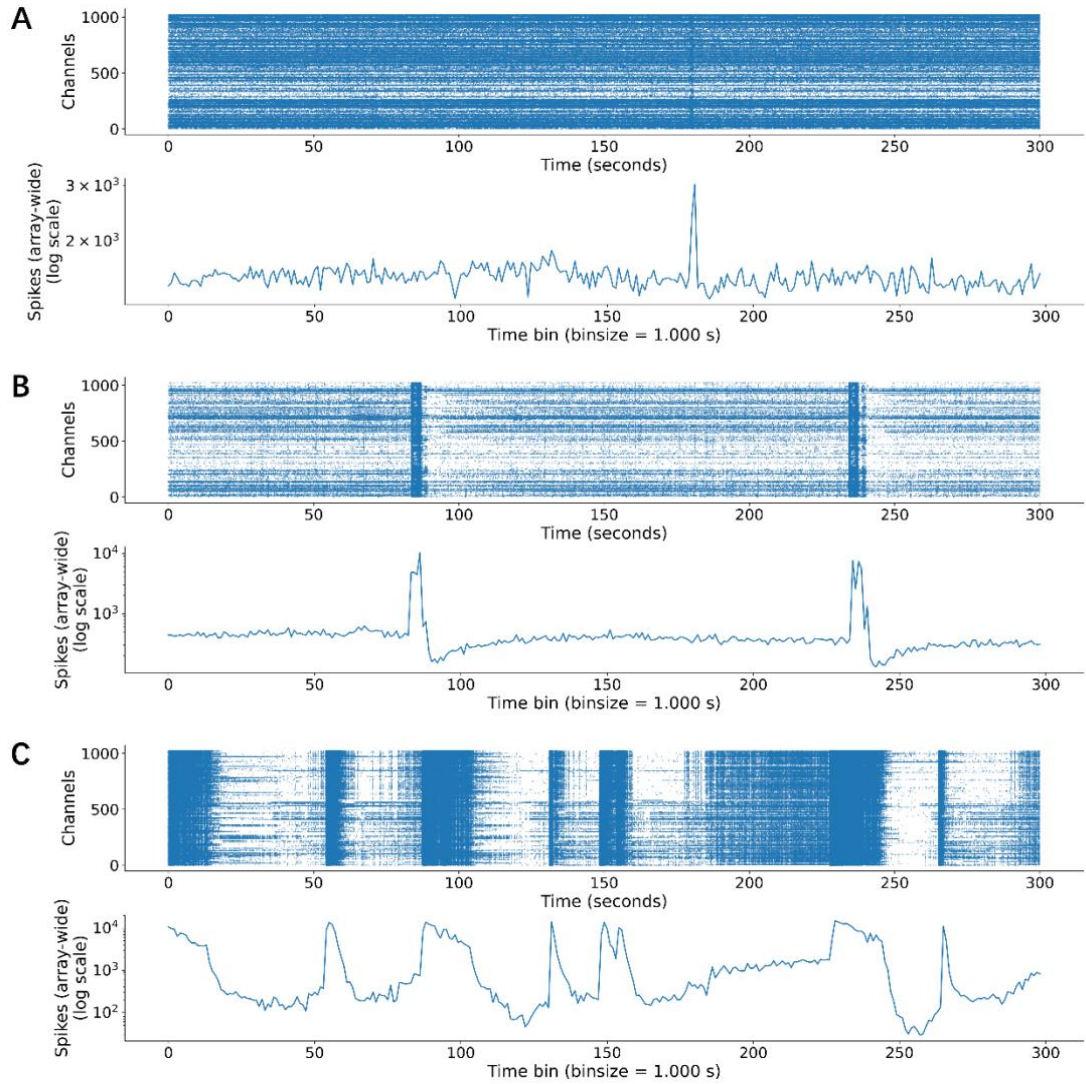

Figure 2. Representative examples of spontaneous activity across avalanche categories. (A) Exponential distribution. Example culture classified as exponential at DIV10. The upper panel shows a 5-min raster plot of spikes recorded from 1,024 channels, and the lower panel shows the corresponding spike counts across all channels as a function of time. (B) Bimodal distribution. Example culture classified as bimodal at DIV10, displayed in the same format as (A). (C) Power-law distribution. Example culture classified as power-law at DIV30, displayed in the same format as (A).

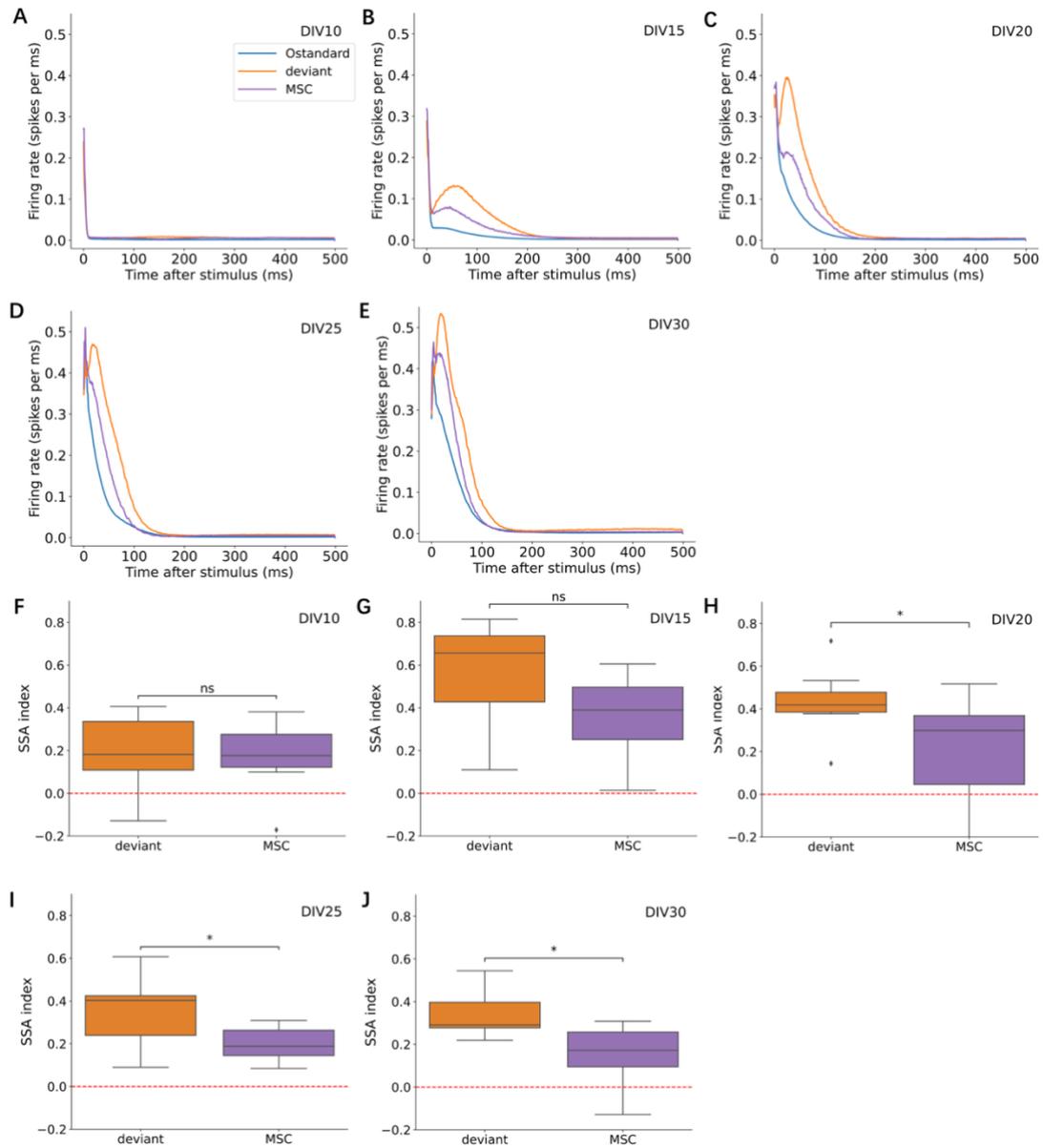

Figure 3. Developmental changes in deviance detection. (A–E) Population peri-stimulus time histograms (PSTHs). Evoked responses at different developmental stages (DIV10, 15, 20, 25, and 30). Late responses (11–150 ms) gradually emerged and increased in amplitude with maturation. (F–J) Stimulus-specific adaptation indices (SSAI). SSAI for deviant and many-standards control (MSC) responses across developmental stages, showing the progressive strengthening of deviance detection.

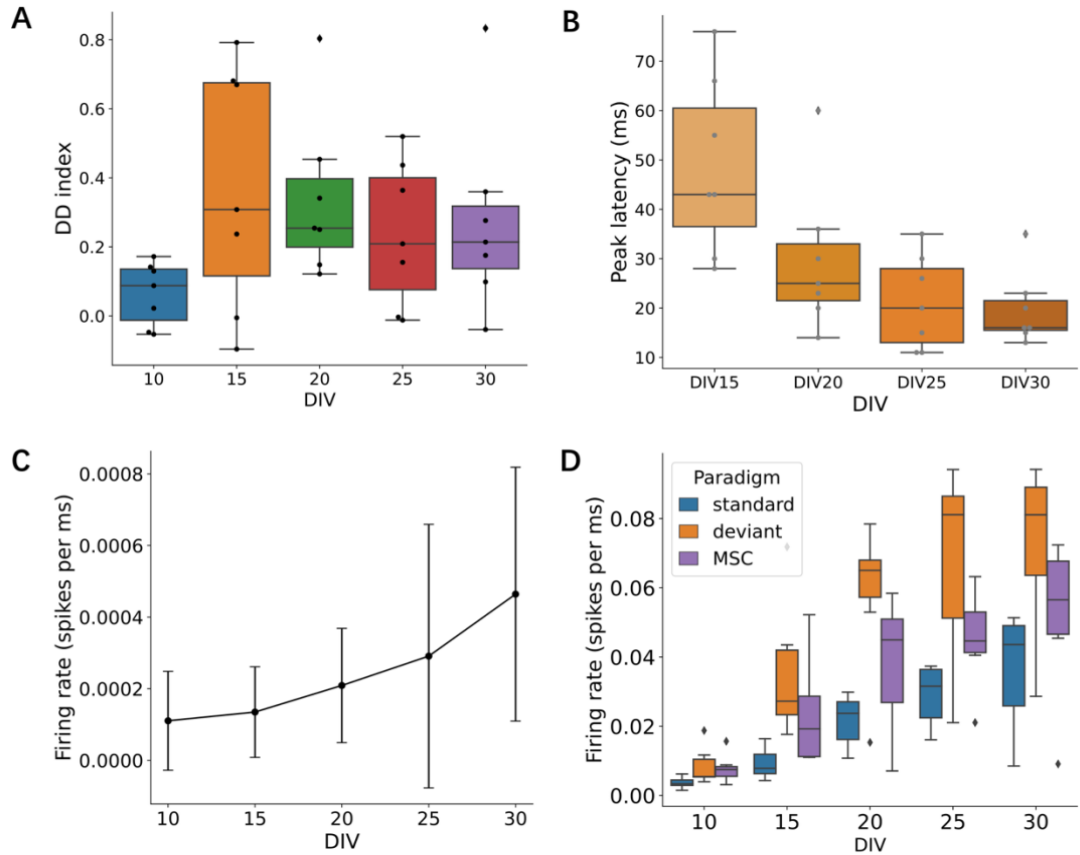

Figure 4. Developmental trajectory of deviance detection and network activity. (A) Deviance detection index (DDI). DDI values across developmental stages (DIV10–30), showing progressive strengthening of deviance detection. (B) Peak latencies. Peak latencies of deviant responses, which progressively shortened with maturation. (C) Spontaneous firing rates. Firing rates increased gradually across development. (D) Evoked response amplitudes. Amplitudes of stimulus-evoked responses, which increased as cultures matured.

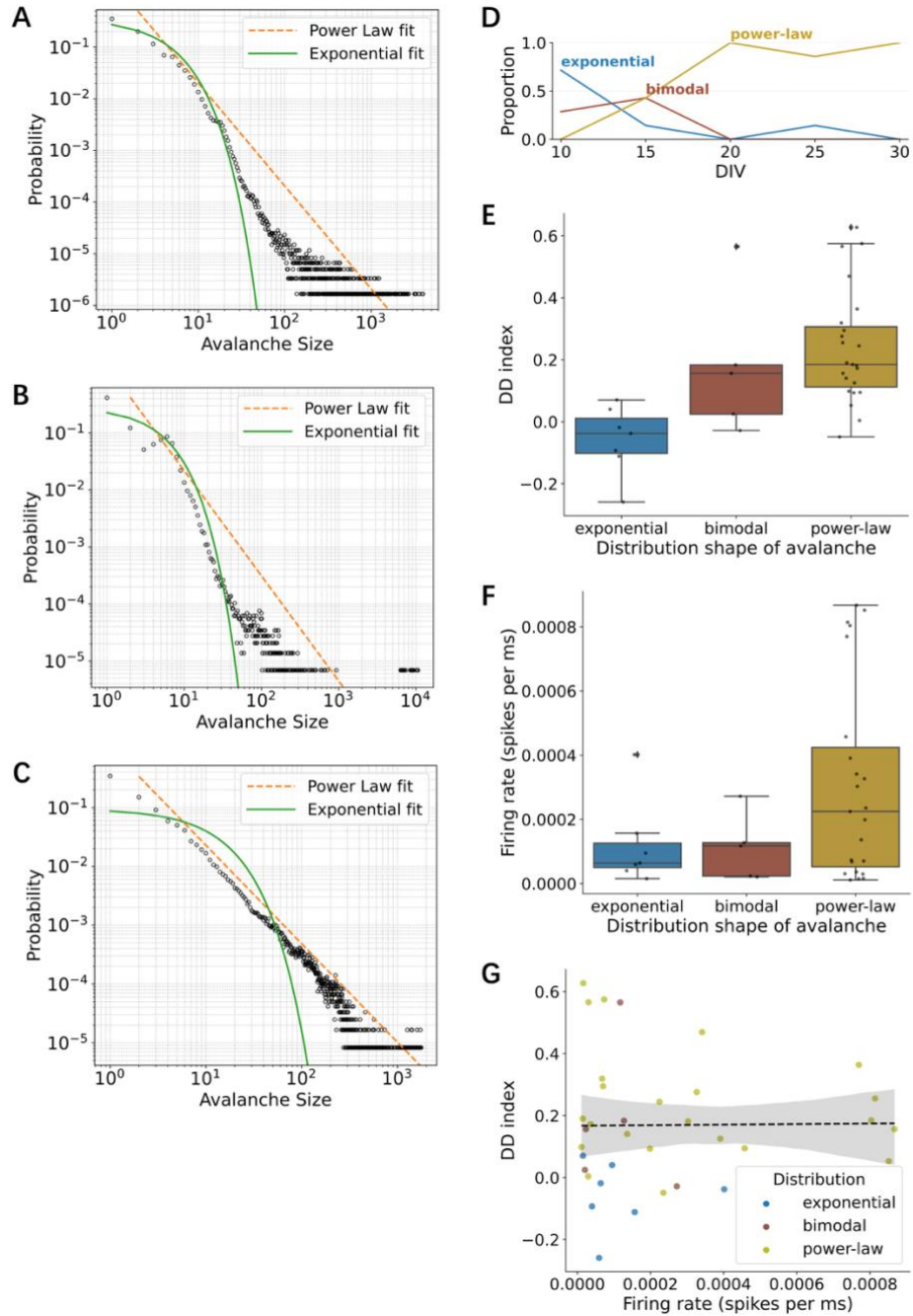

Figure 5. Relationship between avalanche dynamics and deviance detection. (A–C) Representative avalanche size distributions. Examples classified as exponential, bimodal, or power-law. (D) Distribution across developmental stages. Proportion of cultures assigned to each avalanche category over time. (E) Deviance detection indices (DDI). DDI values compared across avalanche categories. Cultures with power-law statistics exhibited significantly stronger DD than those with exponential distributions. (F) Spontaneous firing rates. Firing rates across avalanche categories, showing no significant group differences. (G) DD–firing rate relationship. Scatterplot illustrating no correlation between DDI and firing rate. Error bars and shaded areas represent SEM and 95% confidence intervals, respectively.

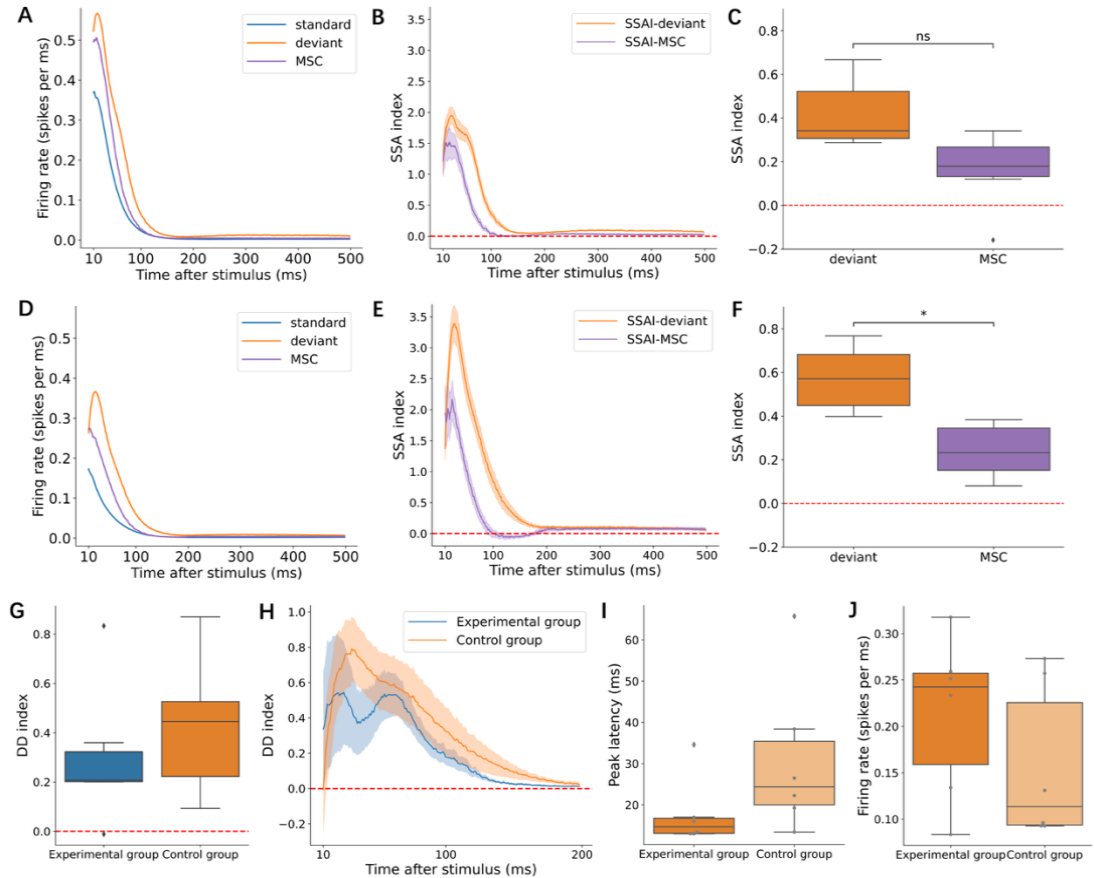

Figure 6. Effects of early stimulation on mature deviance detection. (A–C) Experimental group. Population peri-stimulus time histograms (PSTHs), time-resolved stimulus-specific adaptation (SSAI), and SSAI values for deviant versus MSC stimuli at the mature stage (DIV31–35). (D–F) Control group. Corresponding PSTHs, time-resolved SSAI, and SSAI values. (G) Group comparison of DDI. Deviance detection indices were significantly higher in controls. (H) Time course of DDI. DDI values (10–200 ms) for experimental and control groups. (I) Peak latencies. Deviant peak latencies were shorter in the experimental group. (J) Spontaneous firing rates. Higher firing rates were observed in the experimental group.

## Author Contributions

**Zhuo Zhang:** data curation, formal analysis, investigation, methodology, software, validation, visualization, writing – original draft, writing – review and editing. **Amit Yaron:** investigation, methodology, supervision, validation, writing – review and editing. **Dai Akita:** investigation, methodology, resources, supervision, validation, funding acquisition, writing – review and editing. **Tomoyo Isoguchi Shiramatsu:** methodology, funding acquisition, writing – review and editing. **Zenas C. Chao:** validation, writing – review and editing. **Hirokazu Takahashi:** conceptualization, funding acquisition, project administration, resources, supervision, validation, writing – review and editing.

**Funding**

This work is partly supported by JSPS KAKENHI (23H03465, 24H01544, 24K20854, 25H02600), AMED (24wm0625401h0001), JST (JPMJPR22S8), the Asahi Glass Foundation, and the Secom Science and Technology Foundation.

**Ethics statement**

All animal experimental protocols were approved by the Graduate School of Information Science and Technology's Ethics Committee at the University of Tokyo (JA21-8). All experimental procedures were carried out in accordance with these approved guidelines.

**Conflict of Interest**

The authors declare that the research was conducted in the absence of any commercial or financial relationships that could be construed as potential conflicts of interest.

**Data Availability Statement**

The data that support the findings of this study are available from the corresponding author upon reasonable request.